\documentclass[11pt,a4paper]{article}
\pdfoutput=1
\usepackage{jheppub}
\usepackage{graphicx}

\def\hbar{\hspace{0pt}\raisebox{1pt}{$-$} \hspace{-7pt} h}

\def\5{\overline 5}

\newcommand{\ba}{\begin{eqnarray}}
\newcommand{\ea}{\end{eqnarray}}
\newcommand{\no}{\nonumber}
\newcommand{\be}{\begin{equation}}
\newcommand{\ee}{\end{equation}}
\newcommand{\bea}{\begin{eqnarray}}
\newcommand{\eea}{\end{eqnarray}}

\title{Color $\&$ Weak triplet scalars, the dimuon asymmetry in $B_s$ decay, the top forward-backward asymmetry, and the CDF dijet excess
}
\date{\today
}
\author{
Luca Vecchi}
\affiliation{Theoretical Division T-2, Los Alamos National Laboratory \\ Los Alamos, NM 87545, USA}
\emailAdd{vecchi@lanl.gov}
\abstract{The new physics required to explain the anomalies recently reported by the D0 and CDF collaborations, namely the top forward-backward asymmetry (FBA), the like-sign dimuon charge asymmetry in semileptonic b decay, and the CDF dijet excess, has to feature an amount of \emph{flavor symmetry} in order to satisfy the severe constrains arising from flavor violation. In this paper we show that, once baryon number conservation is imposed, color $\&$ weak triplet scalars with hypercharge $Y=-1/3$ can feature the required flavor structure. 
The color $\&$ weak triplet model can \emph{simultaneously} explain the top FBA and the dimuon charge asymmetry \emph{or} the dimuon charge asymmetry and the CDF dijet excess. However, the CDF dijet excess appears to be incompatible with the top FBA in the minimal framework. Our model for the dimuon asymmetry predicts the observed pattern $h_d\ll h_s$ in the region of parameter space required to explain the top FBA, whereas our model for the CDF dijet anomaly is characterized by the absence of beyond the SM b-quark jets in the excess region. Compatibility of the color $\&$ weak triplet with the electroweak constraints is also discussed. We show that a Higgs boson mass exceeding the LEP bound is typically favored in this scenario, and that both Higgs production and decay can be significantly altered by the triplet. The most promising collider signature is found if the splitting among the components of the triplet is of weak scale magnitude. 
}
\begin{document}
\maketitle
%

\section{Motivations}
\label{intro}

Recently, a number of significant ($\gtrsim3\, \sigma$) anomalies have been reported by the D0 and CDF collaborations. The first is the measurement of a large deviation from the standard model (SM) expectation for the top forward-backward asymmetry~\cite{FBAD0}\cite{FBACDF}; the second is an excess in the $p\overline{p}\to Wjj$ process reported by CDF~\cite{bump}; and the third is an anomaly in the like-sign dimuon charge asymmetry in semileptonic $b$ decay~\cite{like-sign}. While for the first anomaly both collaborations agree that new physics (NP) might play a significant role in explaining the excess, for the second the D0 collaboration has raised serious concerns on the reliability of the CDF analysis~\cite{D0}. Yet, a NP contribution to $Wjj$ in the range of invariant masses for the di-jet system identified by CDF has not been completely excluded by D0. 

A NP explanation of these anomalies requires sizable, flavor-violating couplings to the proton constituents and the third quark generation. This appears to be a challenge in view of the severe bounds imposed on flavor-violation, mainly from the physics of meson oscillation. 

One possibility to extend the SM in a way compatible with flavor constraints is to promote minimal flavor violation~\cite{MFV} to a principle of nature.

Here we emphasize that models in which the NP $\phi$ couples to a di-quark $q_iq_j$ with a coupling $\lambda_{ij}$ antisymmetric in the family indices $i,j$ naturally feature the flavor structure required to explain the Tevatron anomalies. 

Antisymmetry of $\lambda_{ij}$ is automatically ensured if $\phi$ is a color triplet scalar with renormalizable couplings to the SM quarks. There are four possible candidates~\footnote{An incomplete list of references on the physics of colored scalars is given here~\cite{MW}\cite{Gripaios}\cite{list}\cite{Kfactor}\cite{s}.}, which transform under $SU(3)_C\times SU(2)_W$ as $({\bf 3,1})$ and $({\bf 3,3})$. The electro-weak (EW) singlet representations cannot be responsible for the CDF anomaly. Hence, led by a principle of minimality we are left with a single representation of the $SU(3)_C\times SU(2)_W\times U(1)_Y$ group: 
\ba\label{QN}
\phi\sim\left({\bf 3,3}\right)_{-\frac{1}{3}}.
\ea
Our aim is to add to the SM a color $\&$ weak triplet scalar $\phi$ with hypercharge $Y=-1/3$ and discuss the phenomenology of the resulting model. 

If no additional assumption beyond SM gauge invariance is made, the field~(\ref{QN}) triggers proton decay at tree-level via the dimension-6 operator $QQQL/m_\phi^2$. In this paper we will take a phenomenological perspective and assume that the short distance description of our model respects the baryon number. Our approach is somewhat analogous to~\cite{MW}, where the physics of the SM plus a color octet, weak doublet was considered. The main difference between the approach of this latter paper and ours is that in~\cite{MW} minimal flavor violation was assumed while baryon number appeared as an accidental symmetry, whereas here we impose baryon number conservation while make no assumptions on the flavor structure of the theory. In this respect our philosophy is much closer to that of~\cite{Gripaios}.

Color $\&$ weak triplet scalars arise in many extensions of well motivated NP. Our perspective will be however phenomenological, in the sense that we will not assume any of these motivated UV completions. Rather, we will see whether the existence of a weak $\&$ color triplet scalar with the properties required to explain the recent Tevatron anomalies~\cite{FBAD0}\cite{FBACDF}\cite{bump}\cite{like-sign} has anything to teach us about the short distance description of particle physics.

In section~\ref{model} we introduce the color $\&$ weak triplet model. In section~\ref{constraints} we analyze the main phenomenological constraints, arising in particular from flavor violation and the EW bounds. There we will show that antisymmetry of $\lambda_{ij}$ forbids the occurrence of $\Delta F=2$ processes at tree-level; see also~\cite{Gripaios}. One-loop flavor-changing neutral current (FCNC) effects as well as tree-level $\Delta F=1$ processes are found to be consistent with data for \emph{generic} Yukawa couplings as large as $\sim0.1$ and scalars of weak scale masses. In section~\ref{hadron} we discuss resonant production at hadron colliders and the main signatures of the model, and elaborate on possible implications of the triplet on the SM Higgs searches at the LHC. In section~\ref{anomaly} we turn to a discussion of the Tevatron anomalies~\cite{FBAD0}\cite{FBACDF}\cite{bump}\cite{like-sign}. We will argue that the color $\&$ weak triplet model can separately explain all three anomalies. However, a simultaneous explanation can only be provided for two of them, since the CDF $Wjj$ excess appears to be incompatible with the observed top forward-backward asymmetry (FBA). A simultaneous explanation of the CDF dijet excess and the dimuon asymmetry is possible with generic, nonhierarchical couplings, whereas a simultaneous explanation of the top FB asymmetry and the dimuon asymmetry is possible for hierarchical Yukawa matrices. We finally conclude in section~\ref{conclusions}.

\section{The color \& weak triplet model}
\label{model}

Given the quantum numbers in~(\ref{QN}), the most general renormalizable Lagrangian invariant under the baryon number (see section~\ref{intro} for a discussion of our philosophy) is:
\ba\label{C&W}
{\cal L}_{\rm C\& W-triplet}\equiv{\cal L}_{\rm{SM}}+{\rm{Tr}}(D_\mu\phi^\dagger D^\mu\phi)+\lambda_{ij}\overline{Q_i^c}\epsilon\phi Q_j+{\rm{hc}}-V_{\phi^2}-V_{\phi^4}.
\ea
We wrote the color $\&$ weak triplet in matrix notation:
\ba\label{phi}
\phi=
\left( \begin{array}{cc}  \frac{1}{\sqrt{2}}\phi_{-\frac{1}{3}}\quad \phi_{+\frac{2}{3}}  \\
\phi_{-\frac{4}{3}}\quad -\frac{1}{\sqrt{2}}\phi_{-\frac{1}{3}} \end{array}\right),
\ea
where $\phi_{1/3}\equiv\phi_{-1/3}^*,\,\phi_{2/3}\equiv\phi_{-2/3}^*$, and $\phi_{4/3}\equiv\phi_{-4/3}^*$ are complex, $SU(3)_C$ triplet scalars. After electroweak (EW) symmetry breaking these latter become mass eigenstates with electric charges $Q=T^3+Y=1/3,\,2/3,$ and $4/3$, respectively. The covariant derivative reads ${D}_\mu\phi=\partial_\mu\phi+ig_sG_\mu\phi+ig[W_\mu,\phi]+ig'YB_\mu\phi$, where $G_\mu$ is the matrix-valued gluon field and $W_\mu=W^a_\mu\sigma^a/2$ with $\sigma^a$ the Pauli matrices, and the factors of $1/\sqrt{2}$ in~(\ref{phi}) are fixed by the requirement that the components of the weak triplet are canonically normalized.

The interactions with the SM quarks are described by the following Yukawa operator
\ba\label{Yukawa}
{\cal L}_{\rm{Yukawa}}&=&\lambda_{ij}\overline{Q_i^c}\epsilon\phi Q_j+{\rm{hc}}\\\no
&=&\lambda_{ij}\left[-\sqrt{2}\,\overline{u_i^c}\phi_{-\frac{1}{3}}d_j-\overline{d_i^c}\phi_{+\frac{2}{3}}d_j +\overline{u_i^c} \phi_{-\frac{4}{3}}u_j\right]+{\rm{hc.}}
\ea
Here $\epsilon=i\sigma^2$, whereas $Q_i$ are the left handed SM quark doublets and $i,j=1,2,3$ are family indices. The Majorana conjugate of a spinor $\psi$ is defined as usual by $\overline{\psi^c}\equiv\psi^TC$, with $\psi^T$ denoting the transpose and $C=i\gamma^2\gamma^0$ the charge conjugation matrix. For definiteness, we will work in the basis in which the physical fields are related to the gauge eigenstates by $Q_i^T=(V_{ij}u_j, d_i)$, with $V$ the CKM matrix.

The indices $\alpha, \beta, \gamma$ of the fundamental representation of $SU(3)_C$, suppressed in the above expressions, are contracted with the totally antisymmetric tensor $\epsilon^{\alpha\beta\gamma}$. Meanwhile, the $\bf{3}$ of $SU(2)_W$ is symmetric. It follows that the Yukawa operator~(\ref{Yukawa}) is symmetric in the $SU(2)_W$ and spinor indices 
but antisymmetric in the $SU(3)_C$ indices. Hence, gauge invariance requires that:
\ba\label{anti}
\lambda_{ij}=-\lambda_{ji}.
\ea
This property implies the absence of tree-level flavor \emph{diagonal} transitions $i\leftrightarrow i$, and will play a crucial role throughout the paper. See also~\cite{Gripaios}.

The most general quadratic potential for the scalar $\phi$ is
\ba\label{potential}
V_{\phi^2}&=&M_\phi^2\,{\rm{Tr}}(\phi^\dagger\phi)+\kappa_1(H^\dagger H){\rm{Tr}}(\phi^\dagger\phi)+\kappa_2{\rm{Tr}}(\phi^\dagger\sigma^a\phi)H^\dagger\sigma^aH,
\ea
where $H$ is the Higgs doublet and $M_\phi^2,\,\kappa_{1,2}$ are real parameters. 
From~(\ref{potential}) it follows that the tree-level masses of the three components of the triplet are:
\ba\label{tree}\no
m_{1/3}^2&=&M^2_\phi+\kappa_1\frac{v^2}{2}\\
m_{4/3}^2&=&m_{1/3}^2+\kappa_2\frac{v^2}{2}\\\no
m_{2/3}^2&=&m_{1/3}^2-\kappa_2\frac{v^2}{2}=2m_{1/3}^2-m_{4/3}^2,
\ea 
with $v=246$ GeV the SM Higgs vacuum. The pattern~(\ref{tree}) is understood as an effect of isospin violation triggered by $\kappa_2$. We find that
\ba\label{hierarchy}
|\kappa_2|=\frac{2}{v^2}|\Delta m|(2m_{1/3}+|\Delta m|)
\ea
is required to have a mass splitting $|\Delta m|$ among the elements of the weak triplet. Specifically, for $\kappa_2>0$ ($\Delta m<0$) one has $m_{4/3}>m_{1/3}+|\Delta m|$ and $m_{1/3}>m_{2/3}+|\Delta m|$, while for $\kappa_2<0$ ($\Delta m>0$) one has $m_{2/3}>m_{1/3}+\Delta m$ and $m_{1/3}>m_{4/3}+\Delta m$. Plugging in numbers one sees that $|\Delta m|\sim m_W$ is attained for weak scale masses and $|\kappa_2|\sim1$. We will come back to isospin violation when discussing the EW constraints. 

 \section{Experimental constraints}
 \label{constraints}

 \subsection{Flavor violation}

Color $\&$ weak triplets violate flavor at the tree and loop level. In the following we wish to discuss the main constraints arising from $\Delta F=1,2$ processes under the assumption that $\lambda_{ij}$ is a generic, nonhierarchical (antisymmetric) matrix. Hierarchical couplings $|\lambda_{23}|\ll|\lambda_{12}|\ll|\lambda_{13}|\sim1$ will be discussed in section~\ref{hierarchical}.


\subsubsection{$\Delta F=2$ processes}

Integrating out $\phi$ at tree level, and after a Fierz transformation, we find the effective Lagrangian 
\ba\label{DF=1}
{\cal L}^{\rm{Tree}}_{\Delta F=1}&=&\frac{\lambda_{ij}\lambda^*_{kl}}{2m_{4/3}^2}\left[(\bar u_{k}^\alpha\gamma^\mu u_i^\alpha)(\bar u_{l}^\beta\gamma_\mu u_j^\beta)-(\bar u_{k}^\beta\gamma^\mu u_i^\alpha)(\bar u_{l}^\alpha\gamma_\mu u_j^\beta)\right]+{\rm hc}\\\no
&+&\frac{\lambda_{ij}\lambda^*_{kl}}{2m_{2/3}^2}\left[(\bar d_{k}^\alpha\gamma^\mu d_i^\alpha)(\bar d_{l}^\beta\gamma_\mu d_j^\beta)-(\bar d_{k}^\beta\gamma^\mu d_i^\alpha)(\bar d_{l}^\alpha\gamma_\mu d_j^\beta)\right]+{\rm hc}\\\no
&+&\frac{\lambda_{ij}\lambda^*_{kl}}{m_{1/3}^2}\left[(\bar u_{k}^\alpha\gamma^\mu u_i^\alpha)(\bar d_{l}^\beta\gamma_\mu d_j^\beta)-(\bar u_{k}^\beta\gamma^\mu u_i^\alpha)(\bar d_{l}^\alpha\gamma_\mu d_j^\beta)\right]+{\rm hc},
\ea
where $\alpha, \beta$ and $i,j,k,l$ are color and flavor indices, respectively. Throughout the text all spinors are left handed, so for brevity we will always write $q_i$ to mean its left-handed component. 

The operators~(\ref{DF=1}) do not lead to $\Delta F=2$ transitions. Indeed, it is easy to see that from the antisymmetry of $\lambda_{ij}$ it follows that all $\Delta F=2$ operators in~(\ref{DF=1}) vanish. 

The absence of tree-level $\Delta F=2$ processes can also be understood using the spurion approach. In this language the matrix $\lambda$ has two antisymmetric indices with values in the $SU(3)_Q$ family group, and, without mass insertions,  $\Delta F=2$ operators of the form $(\overline{Q_i}\gamma^\mu Q_j)^2$ require at least \emph{four} powers of $\lambda$ with the structure $(\sum_{n}\lambda_{jn}\lambda^*_{in})^2$. This structure cannot clearly be generated at tree-level, but will be generated at one-loop level, as we will see shortly. 


We here focus on $K-\overline{K}$ mixing because this physics provides the strongest constraint on NP contributions to $\Delta F=2$ processes if $\lambda_{ij}$ is nonhierarchical. The situation changes when hierarchical Yukawa matrices are considered, see section~\ref{hierarchical}. 

The relevant operator in the present analysis is
\ba\label{KK}
C^1_K(\overline{d}^\alpha\gamma^\mu s^\alpha)(\overline{d}^\beta\gamma^\mu s^\beta).
\ea
The operator~(\ref{KK}) arises from 2 box diagrams with internal $\phi_{1/3}$ and 2 box diagrams with internal $\phi_{2/3}$ boson lines, while one-loop diagrams involving a $W^\pm$ and a $\phi$ vanish, see also~\cite{s}. This is understood by noting that the color structure of the diagrams with a single $\phi$ arises from two powers of the totally antisymmetric color tensor (i.e. two~(\ref{Yukawa}) vertices) and two powers of the Kronecker delta function (i.e. one $\phi$ and one $W$ propagators) and is therefore of the form $\epsilon^{\alpha\beta\gamma}\epsilon^{\dot\alpha\dot\beta\dot\gamma}\delta^{\alpha\dot\alpha}(\overline{d^{\dot\beta}}\gamma^\mu s^\beta\overline{d^{\dot\gamma}}\gamma_\mu s^\gamma)=(\delta^{\beta\dot\beta}\delta^{\gamma\dot\gamma}-\delta^{\beta\dot\gamma}\delta^{\gamma\dot\beta})(\overline{d^{\dot\beta}}\gamma^\mu s^\beta\overline{d^{\dot\gamma}}\gamma_\mu s^\gamma)$, which identically vanishes after a Fierz transformation. This is also in accord with our expectation from the spurion argument that $\Delta F=2$ processes must involve at least four powers of $\lambda$. The color structure of the diagrams with two virtual $\phi$'s instead arises from four powers of the totally antisymmetric color tensor (i.e. four ~(\ref{Yukawa}) vertices) and two powers of the Kronecker delta function. This will lead to the expected $\Delta F=2$ Yukawa structure.

The Wilson coefficient of~(\ref{KK}) evaluated at a scale $m_\phi=O(m_{1/3},m_{2/3})$ is 
\ba\label{C1}
C^1_K(m_\phi)&=&\sum_{j,j'}\frac{\xi_j\,\xi_{j'}}{16\pi^2m_{1/3}^2}\,\,\,{\cal I}(x^u_j,x^u_{j'})+\sum_{j,j'}\frac{\xi_j\xi_{j'}}{4\pi^2m_{2/3}^2}\,\,\,{\cal I}(x^d_3,x_{3}^d)\\\no
\xi_j&=&\lambda_{2i}V_{ij}V^*_{kj}\lambda^*_{1k}.
\ea
In eqs.~(\ref{C1}) we defined $x^{u}_i=m_{u_i}^2/m_{1/3}^2$ and $x^{d}_i=m_{d_i}^2/m_{2/3}^2$. 
Note that $\sum_i\xi_i=\sum_i\lambda_{2i}\lambda^*_{1i}=\lambda_{23}\lambda^*_{13}$, where in the first equality we used the unitarity of the CKM matrix and in the second the antisymmetry of $\lambda$. The loop function
\ba
{\cal I}(x,y)=\frac{x^2\log x}{(1-x)^2(x-y)}-\frac{y^2\log y}{(1-y)^2(x-y)}+\frac{1}{(1-x)(1-y)}
\ea
is normalized such that ${\cal I}(0,0)=1$.

The differences between the $\phi_{1/3}$ and $\phi_{2/3}$ contributions stem from the enhancement factor $(\sqrt{2})^4$ in the $\phi_{2/3}$ coupling, see~(\ref{Yukawa})~\footnote{Note that $\sum_{i,j}\lambda_{ij}q^\alpha_iq^\beta_j\epsilon_{\alpha\beta\gamma}=2(\lambda_{12}q^\alpha_1q^\beta_2+\lambda_{23}q^\alpha_2q^\beta_3+\lambda_{13}q^\alpha_1q^\beta_3)\epsilon_{\alpha\beta\gamma}$}, the different mass, and the fact that by antisymmetry of $\lambda$ the $\phi_{2/3}$ diagrams can only involve virtual $b$-quarks.

If the SM fermions were degenerate in mass we would have ${\cal I}(x_{j},x_{j'})=const$, and the only source of flavor violation would come from $\lambda$. Using the unitarity of the CKM matrix one sees that~(\ref{C1}) would then scale as $(\lambda_{23}\lambda^*_{13})^2$, which is the structure anticipated by the spurion analysis. 

More generally, since we focus on models with $m_\phi$ of the order the weak scale, neglecting the SM quark masses with the exception of the top mass is always a good approximation. Using unitarity of the CKM matrix and the symmetry property ${\cal I}(x,y)={\cal I}(y,x)$, and taking $x_1^u,\,x^u_2,\,x^d_3\ll1$, we write
\ba\label{a}
\xi_j\,\xi_{j'}\,\,{\cal I}(x^u_j,x^u_{j'})&\approx&(\sum_i\xi_i)^2{\cal I}(0,0)\\\no
&+&2\xi_3(\sum_i\xi_i)\left[{\cal I}(x^u_3,0)-{\cal I}(0,0)\right]\\\no
&+&\xi_3^2\left[{\cal I}(0,0)+{\cal I}(x^u_3,x^u_3)-2{\cal I}(x^u_3,0)\right].
\ea
Eq.~(\ref{a}) can be studied in two limits. The first applies to non-hierarchical, antisymmetric matrices $\lambda$, for which we find $\sum_i\xi_i\approx\xi_3$. The second limit applies when the couplings $\lambda_{13,23}$ to the third SM quark generations are suppressed compared to $\lambda_{12}$, for which $\sum_i\xi_i\ll\xi_3$. We will study the latter limit in section~\ref{hierarchical}.

\begin{figure}
\begin{center}
\includegraphics[width=4.0in]{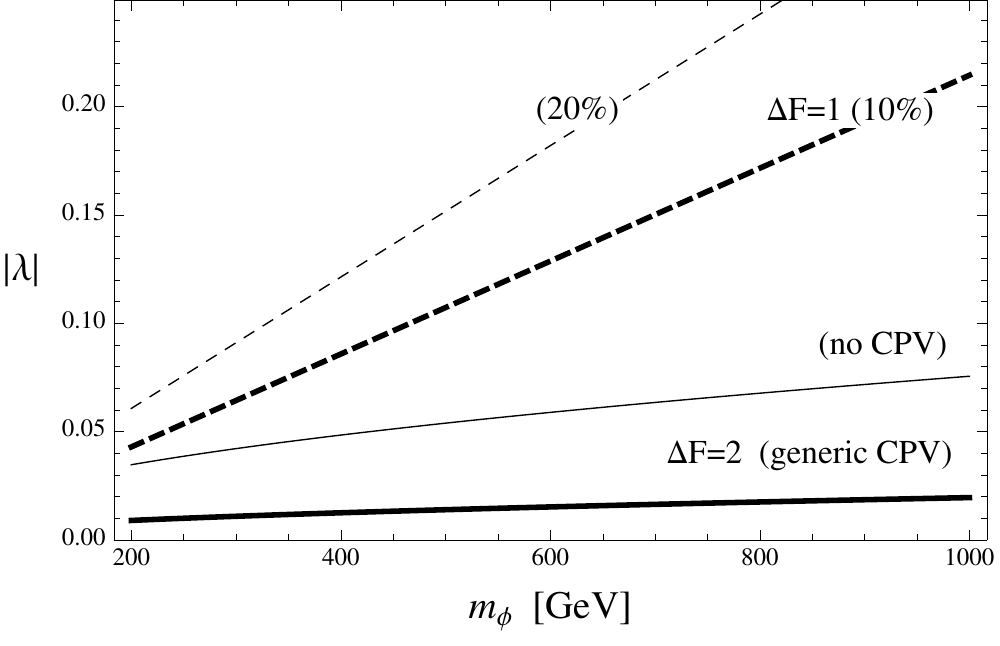}
\caption{\small (Nonhierarchical Yukawa couplings) Bounds from $K-\overline{K}$ mixing (solid curves labeled by ``$\Delta F=2$") and $B\rightarrow J/\Psi K, \phi_s K, X_s\gamma$ (dashed curves labeled by ``$\Delta F=1$") on $|\lambda|\equiv\sqrt{|\lambda_{23}\lambda^*_{13}|}$ and $|\lambda|\equiv\sqrt{|\lambda_{12}\lambda_{13}^*|}$, respectively, as a function of $m_{2/3}=m_{1/3}$ (see text). In presenting the $\Delta F=2$ bounds we distinguished between scenarios with $O(1)$ phases (thick, solid line labeled by ``generic CPV'') and vanishing phases (thin, solid line labeled by ``no CPV''). The $\Delta F=1$ constraints are derived by requiring that the NP corrections induced by the color $\&$ weak triplet do not exceed $10\%$ (thick-dashed line) or $20\%$ (thin-dashed line) of the SM prediction. The bound arising from $B\to X_s\gamma$ essentially overlaps with the $20\%$ line. A mild, generic suppression of $\lambda_{13,23}$ compared to the coupling $\lambda_{12}$ to the first two generations allows $|\lambda_{12}|=0.05-0.1$ (see text). Hierarchical couplings will be discussed in section~\ref{hierarchical}.
\label{FCNC}}
\end{center}
\end{figure}

For generic matrices $\lambda$, in which case $\lambda_{12,13,23}$ have comparable magnitude, eq.~(\ref{a}) is well approximated by taking the limit $V_{ij}\rightarrow\delta_{ij}$. In this case $\xi_3\rightarrow\lambda_{23}\lambda^*_{13}=\sum_i\xi_i$, and the Wilson coefficient~(\ref{C1}) becomes
\ba\label{C1app}
C^1_K(m_\phi)\approx\left(\frac{\sum_i\xi_i}{2\pi}\right)^2\left[\frac{{\cal I}(x^u_3,x^u_3)}{4m_{1/3}^2}+\frac{{\cal I}(0,0)}{m_{2/3}^2}\right]~~~~~~~~~~~~~~~~({\rm{generic~}}\lambda_{ij}).
\ea
Assuming $O(1)$ CP-violating phases we impose the $95\%$ CL bound~\cite{UTFit}
\ba\label{boundKK}
|C_K^1(m_{\phi})|\lesssim\left(\frac{1}{1.5\times10^4\,\,{\rm{TeV}}}\right)^2.
\ea
The constraint on the NP scale relaxes to $10^3$ TeV if CP violation in the NP sector is suppressed~\cite{UTFit}. We plot the resulting bounds on $\sqrt{|\lambda_{23}\lambda^*_{13}|}$ in fig.~\ref{FCNC} for $m_{2/3}=m_{1/3}$. In the presence of mass splitting the bound is controlled by the $\phi_{2/3}$ loops, and is essentially unaffected for $|\Delta m|\lesssim100$ GeV. 

Analogous expressions can be derived for $D-\overline{D}$ or $B_q-\overline{B}_q$ mixing. For generic Yukawa matrices the relevant couplings are $\sqrt{|\lambda_{13}\lambda_{23}^*|}$ for the $1\leftrightarrow 2$ transitions and $\sqrt{|\lambda_{32}\lambda_{12}^*|}$ for the $1\leftrightarrow 3$ or $\sqrt{|\lambda_{21}\lambda_{31}^*|}$ for the $2\leftrightarrow 3$ processes. The bounds arising from $D-\overline{D}$ or $B_q-\overline{B}_q$ mixing are always weaker than those shown in fig.~\ref{FCNC} when the couplings are nonhierarchical. This conclusion changes when hierarchical matrices $\lambda_{ij}$ are assumed, see section~\ref{hierarchical}.

\subsubsection{$\Delta F=1$ processes}
\label{DFu1}

The effective Lagrangian~(\ref{DF=1}) contributes to non-leptonic, $\Delta F=1$ decays of $B$ or $K$ mesons (mediated by $\phi_{1/3}$ and $\phi_{2/3}$) and $D$ mesons (mediated by $\phi_{1/3}$ and $\phi_{4/3}$). We find that the most stringent bounds are on the coefficients of the operators associated to $B$-meson decay. The relevant operators can be divided into two classes. The first class of operators has the same current-current, 4-flavors structure as the one generated at tree-level in the SM. The second class includes the penguin operators $(\overline{b}s)_{V-A}(\overline{q}q)_{V-A}$. At the scale $\mu=m_W$ these are generated within the SM by QCD and EW penguin diagrams. The second class leads to stronger bounds than the first class, so we will focus on it in what follows. 
We will consider for definiteness the $|\Delta B|=|\Delta S|=1$ transitions. Comparable (weaker) bounds are expected to arise from other nonleptonic $\Delta B=1$ processes. 

Neglecting the small off-diagonal components of the CKM matrix, the $|\Delta B|=|\Delta S|=1$ operators derived from~(\ref{DF=1}) are
\ba\label{DB}
{\cal L}_{|\Delta B|=|\Delta S|=1}^{\rm Tree}&=&\frac{\lambda_{12}\lambda^*_{13}}{m_{1/3}^2}\left[(\bar b^\alpha\gamma^\mu s^\alpha)(\bar u^\beta\gamma_\mu u^\beta)-(\bar b^\beta\gamma^\mu s^\alpha)(\bar u^\alpha\gamma_\mu u^\beta)\right]+{\rm hc}\\\no
&+&\frac{\lambda_{12}\lambda^*_{13}}{m_{2/3}^2}\left[(\bar b^\alpha\gamma^\mu s^\alpha)(\bar d^\beta\gamma_\mu d^\beta)-(\bar b^\beta\gamma^\mu s^\alpha)(\bar d^\alpha\gamma_\mu d^\beta)\right]+{\rm hc},
\ea
where as usual all spinors are left-handed (note that the factor of $2$ suppressing the $\phi_{2/3}$ operators in~(\ref{DF=1}) has disappeared as a consequence of (approximate) isospin invariance). 

NP contributions from the nonstandard operators~(\ref{DB}) to $B\rightarrow\pi K$ decays were derived in~\cite{Grossman}. The strongest constraint  on our effective field theory~(\ref{DB}) arises in the case of large CP-violation and reads $\sqrt{|\lambda_{12}\lambda_{13}^*|}\lesssim0.5\,(m_{\phi}/1\,{\rm{TeV}})$, see eq.(3.10) of~\cite{Grossman}.

More stringent bounds arise from the $B\rightarrow J/\Psi K$ and $B\rightarrow\phi_s K$ processes~\cite{Lunghi}~\footnote{ Here $\phi_s$ denotes the QCD $s\overline{s}$ meson!}. In order to conform with the conventions of~\cite{Lunghi} we assume that the splitting among the components of the triplet vanishes. Under this simplifying assumption we can work with the SM effective operator basis
\ba
\frac{4G_F}{\sqrt{2}}V_{33}V^*_{32}\,\sum_{n=1}^{10+2}C_{n}(\mu){\cal O}_n,
\ea
and find that the operators that are corrected by tree-level exchange of the color $\&$ weak triplet are the QCD penguin operators: 
\ba\label{btos}
{\cal O}_3=\overline{s}^\alpha \gamma^\mu b^\alpha\sum_q \overline{q}^\beta\gamma_\mu q^\beta~~~~~~&&~~~~~~~~{\cal O}_4=\overline{s}^\alpha \gamma^\mu b^\beta \sum_q \overline{q}^\beta \gamma_\mu q^\alpha,
\ea
where $q=u,d,s,c,b$ and again all fields are left-handed. 
Within our approximations $q=u,d$, see eq.(\ref{DB}). However, for simplicity we will conservatively assume that the NP corrections are universal in $q$ and equal to the largest contribution. Again ignoring the RG flow from $\mu=m_{1/3}=m_{2/3}$ down to $\mu=m_W$, we find
\ba\label{delta}
\frac{4G_F}{\sqrt{2}}V_{33}V^*_{32}\,\delta C_{3}(m_W)=\frac{\lambda_{13}\lambda_{12}^*}{m_\phi^2}~~~~~~~~~~~~~\delta C_{4}(m_W)=-\,\delta C_{3}(m_W).
\ea
where we defined the Wilson coefficients as a sum $C_n=C_n^{SM}+\delta C_n$ of the SM plus NP contributions. Using the results of~\cite{Lunghi} (see their eqs.(20) and (21)) we write the total (SM + NP) expression for the parameters $C_{\phi_s}$ and $C_\Psi$ controlling $B\rightarrow\phi_s K$ and $B\rightarrow J/\Psi K$ respectively as:
\ba\label{CP}
C_{\phi_s}(m_b)&=&-0.0243\left(1.0710-0.0050\frac{\delta C_4(m_W)}{C_4^{SM}(m_W)}\right)\\\no
C_{\Psi}(m_b)&=&+0.3515\left(1.0105+0.0040\frac{\delta C_4(m_W)}{C_4^{SM}(m_W)}\right),
\ea
where $\delta C_4(m_W)$ is given in~(\ref{delta}), while the SM term reads 
\ba
C_4^{SM}(m_W)=-3\,C_3^{SM}(m_W) =\left(E(m_t^2/m_W^2)-\frac{2}{3}\right)\frac{\alpha_s(m_W)}{8\pi}.
\ea
Here $E(x)$ is a standard one-loop function (see for example~\cite{btos} after having switched to our basis~(\ref{btos})).

We require that the NP contribution to $C_{\phi_s}$ and $C_\Psi$ be at most $10\%$ ($20\%$) of the SM prediction. It is important to stress that this bound is somewhat arbitrary given the large uncertainty associated to the SM predictions for these processes. 

An inspection of~(\ref{CP}) shows that the bound arising from $B\rightarrow\phi_s K$ is slightly stronger. We plot the resulting constraint on $\sqrt{|\lambda_{13}\lambda_{12}^*|}$ in fig.~\ref{FCNC} together with the bound from $K$-meson oscillation (which refers to $\sqrt{|\lambda_{23}\lambda^*_{13}|}$). Given the large suppression of the NP correction in~(\ref{CP}) we find that $|\delta C_4(m_W)/C_4^{SM}(m_W)|=O(10)$ is allowed~\cite{Lunghi}. As a result, the $\Delta F=1$ constraints are always weaker than those from meson oscillation.

We expect our $\Delta F=1$ bounds not to change quantitatively -- actually to relax a bit -- in the presence of a mass splitting among the components of the triplet. The physics would however be qualitatively different in that case since there would be potentially sizable corrections to the EW penguin operators. It would be interesting to analyze the effect of these corrections in light of the apparent inconsistencies between theory and experiments in the $B\rightarrow J/\Psi\phi_s$ system reported in~\cite{JPsiPhi}.

Next we study the NP effects on $B\rightarrow X_s\gamma$ transitions. These are essentially controlled by the penguin operator
\ba
{\cal O}_7=\frac{em_b}{16\pi^2}\overline{s}_L\sigma_{\mu\nu}b_RF^{\mu\nu}.
\ea
Integrating out the triplet we find that the dominant NP contribution to $C_7$ comes from $\phi_{2/3}$ and approximately reads $|\delta C_7|\approx2|\lambda_{31}\lambda_{21}^*|v^2/(3|V_{32}^*V_{33}|m_\phi^2)$, where, following our convention, we defined $C_7=C_7^{\rm SM}+\delta C_7$. We conservatively require $|\delta C_7|\lesssim0.1$ (see for instance~\cite{Isidori}), and find that 
\ba\label{BtoS}
\sqrt{|\lambda_{12}\lambda_{13}^*|}\lesssim0.3\,(m_{\phi}/1\,{\rm{TeV}}). 
\ea
This constraint essentially coincides with the $20\%$ bound shown in fig.~\ref{FCNC}. 

Flavor-conserving dipole moment operators 
are not severely constrained in our model~\cite{Gripaios}, while the $K\to\pi\nu\overline{\nu}$ decays are found to be under control for generic couplings satisfying the ``no CPV" bound in the figure~\cite{ratt}.

Before turning to a study of the EW constraints on the color $\&$ weak triplet scalar model, we find it worthwhile to comment the bounds presented in fig.~\ref{FCNC}.

If one assumes that $\lambda_{12,13,23}$ are comparable in magnitude, fig.~\ref{FCNC} implies that $|\lambda_{ij}|\lesssim0.01-0.02$, as required by Kaon physics. However, $K-\overline{K}$ must involve both $\lambda_{13}$ and $\lambda_{23}$ to be effective, and hence the bounds on the couplings to the first and second generations can generally be relaxed if either $\lambda_{13}$ or $\lambda_{23}$ are somewhat suppressed. An interesting limit is the one in which $|\lambda_{13}|$ and $|\lambda_{23}|$ are suppressed by $O(1)$ numbers compared to $|\lambda_{12}|$, in which case all the bounds shown in fig.~\ref{FCNC} can be satisfied with  
\ba\label{generic}
|\lambda_{12}|\lesssim0.05-0.10~~~~~~~~~~({\rm generic}~|\lambda_{ij}|).
\ea
For example, if we take $|\lambda_{13}|=|\lambda_{23}|=a|\lambda_{12}|$, with $a$ a positive number, and assume $O(1)$ CP violating phases, it is sufficient to have $a=0.2$ to allow $|\lambda_{12}|=0.05$, while $a=0.1$ to allow $|\lambda_{12}|=0.10$.

\subsection{Electroweak constraints}
\label{EWbounds}

At one-loop, the EW $S$ and $T$ parameters for an EW triplet are:
\ba\label{ST}
S&=&-\frac{YN_c}{3\pi}\log\frac{m_{2/3}^2}{m_{4/3}^2}\\\no
T&=&\frac{N_c}{4\pi\sin^2\theta_w}\frac{A(m_{1/3},m_{4/3})+A(m_{1/3},m_{2/3})}{m_W^2}, 
\ea
with
\ba
A(x,y)=\frac{x^2+y^2}{2}-\frac{x^2y^2}{x^2-y^2}\log\frac{x^2}{y^2}.
\ea
We derived them using the general expressions given in the Appendix of~\cite{Barbieri}, see also~\cite{T}. In our model $N_c=3$ and $Y=-1/3$. Note that $T$ is positive definite while $S$ can be either positive or negative.

In writing $S$ we assumed that $[\Pi_{Y3}(m_Z)-\Pi_{Y3}(0)]/m_Z\approx\Pi_{Y3}'(0)$. This corresponds to neglecting a small, negative remainder 
on the right hand side of the first equation in~(\ref{ST}), see~\cite{T}. 
For $m_\phi\gtrsim100$ GeV the correction is $O(10^{-2})$ and can be safely neglected in what follows. 

We will also neglect the other EW parameters $U, V, W, X$ since we find that these are negligible compared to $S,T$. This is expected to be a conservative assumption because the allowed region for $U=0$ is slightly reduced compared to $U\neq0$~\cite{STfit}.

The contribution to the $T$ parameter is enhanced compared to the case of a weak doublet, color singlet representation, and hence isospin violation is severely constrained in this model. For example, if the triplet is assumed to saturate the NP contribution to the EW precision parameters, the mass splitting $\Delta m\equiv m_{1/3}-m_{4/3} $ is bounded at the $99\%$ C.L. to lie in the regime $|\Delta m|\lesssim50-60$ GeV by the $T$ parameter alone (assuming no correlation with $S,U,\dots$) when $m_{\phi}\gtrsim200$ GeV and for perturbative Higgs masses $m_h\lesssim1000$ GeV.

\begin{figure}
\begin{center}
\includegraphics[width=3.5in]{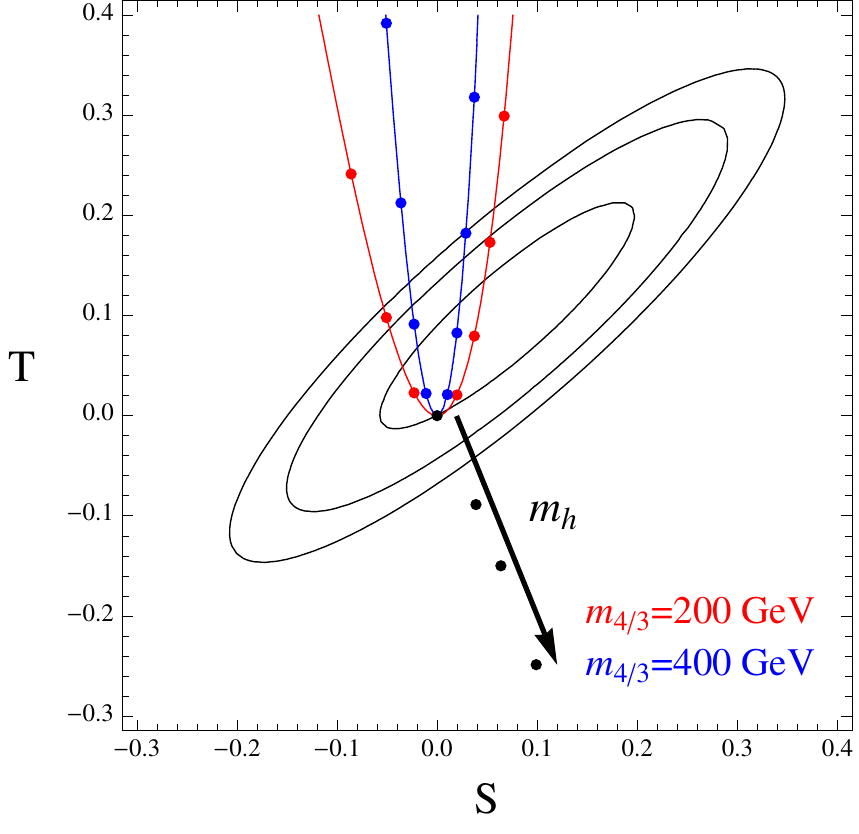}
\caption{\small $68\%$, $95\%$, and $99\%$ C.L. allowed region in the $S-T$ plane ($U=0$) as taken from~\cite{STfit}. Here we assume that no NP contributions to the EW parameters besides those from $\phi$ are present, and that eqs.~(\ref{tree}) are satisfied. We used the exact one-loop expressions~(\ref{ST}). The red (blue) lines and dots are for $m_{4/3}=200\, (400)$ GeV and a SM Higgs mass $m_h=120$ GeV. They refer, from left top to right top, to the different choices $m_{1/3}-m_{4/3}=-40, -30, -20, -10, 0, +10, +20, +30, +40$. The origin $m_{1/3}-m_{4/3}=0$ of the red and blue curves (and with it the curves themselves, rigidly) shifts as indicated by the black arrow if the SM Higgs mass is increased from the reference value to $m_h=300,\, 500,$ and $1000$ GeV. For $m_{4/3}\gtrsim500$ GeV the $S,T$ parameters are well approximated by eqs.~(\ref{STapp}).\label{LEP-ST}}
\end{center}
\end{figure}

In fig.~\ref{LEP-ST} we show the EW constraints in the $S-T$ plane for $m_{4/3}=200,\, 400$ GeV and various values of $m_{1/3}$. In the figure, $m_{2/3}$ is determined by eq.~(\ref{tree}) and the experimental ellipses are from~\cite{STfit}. 
Compatibility with experiments is improved as the SM Higgs mass is increased above the reference value $m_h=120$ GeV adopted in~\cite{STfit}. SM Higgs masses of several hundred GeV, or even $m_h=O($TeV$)$, are preferred as long as $\Delta m\neq0$. 

For larger masses ($m_{4/3}>400$ GeV) and small splitting ($|\Delta m|<50$ GeV) eqs.~(\ref{ST}) become, to a high accuracy,  
\ba\label{STapp}
S\approx-\frac{4}{3\pi}YN_c\frac{\Delta m}{m_\phi}~~~~~~~~~~~~T&\approx& \frac{N_c}{6\pi\sin^2\theta_w}\left(\frac{\Delta m}{m_W}\right)^2
\ea
where, at leading order, $m_\phi$ is any of $m_{1/3}$, $m_{2/3}$, or $m_{4/3}$. We see that in the large mass regime the $T$ parameter scales with $\Delta m$ such that $T\propto S^2$, which explains the origin of the approximate parabolas visible in fig.~\ref{LEP-ST}.

We will see that the ideal scenario for the detection of the color $\&$ weak triplet is when $|\Delta m|\gtrsim m_W$, so it is worth to comment on this latter possibility. In order to accommodate a splitting $|\Delta m|>60$ GeV, the model requires additional, \emph{negative} NP contributions to $T$ or a violation of the tree-level relation~(\ref{tree}). 

Violations of~(\ref{tree}) are expected to arise at the quantum level. However, in a perturbative description this effect is  small, and we therefore ignore this possibility. 

For what concerns the former possibility, two simple examples come to mind. In the first example an EW triplet, color neutral scalar with $Y=2$ is added to the theory. This way a negative contribution to $T$ arises at tree-level if the neutral component of the new triplet gets a vacuum expectation value. The color $\&$ weak triplet model can now easily be made consistent with the EW constraints for a wide spectrum of $|\Delta m|$'s by adjusting the vacuum of the new scalar.

In the second example an EW doublet scalar with the same quantum numbers of the SM Higgs is added. In this case one finds that the one-loop contribution to $T$ of the axial ($A$) and CP-even ($H$) scalars is negative. Assuming for simplicity that this new field does not participate to EW symmetry breaking, and using the results of~\cite{Barbieri}, we see that $(S_{tot}, T_{tot})=(T+T_{doublet}, S+S_{doublet})$ can lie within the 2 standard deviations ellipse of fig.~\ref{LEP-ST} even for $|\Delta m|=$ few $\times\, 100$ GeV. For example, fixing $m_h$ at its reference value and $m_{4/3}=150$ GeV and $m_{1/3}=300$ GeV -- and hence $m_{2/3}\approx397$ GeV from~(\ref{tree}) -- we find that $(S_{tot},T_{tot})\approx(0.21,0.27)$ when $m_H=570$ GeV, $m_A=1430$ GeV, and $m_{H^\pm}=1000$ GeV.

The message to be taken from these two examples is that one cannot exclude scenarios with $|\Delta m|$ as large as the weak scale on the basis of EW precision data alone, as new ingredients besides the triplet $\phi$ might be present in a complete theory. We discussed the cases in which this new ingredient is either an additional EW triplet, color singlet scalar or an EW doublet scalar, but many alternative scenarios are possible.

\section{The triplet at hadron colliders}
\label{hadron}

\subsection{Production and decay}
\label{resprod}

Gauge and Lorentz invariance forbid the processes $gg,qg\rightarrow\phi$, so the production of the triplet $\phi$ at hadron colliders dominantly proceeds via $qq$ fusion. In this section we will discuss $\phi_{1/3}$ resonant production. Similar production cross sections are expected for $\phi_{2/3,4/3}$.

At leading order, the cross section $\sigma(p\bar p/pp\rightarrow\phi_{1/3})$ at a proton-anti proton/proton-proton collider is given by
\ba\label{LO}
\sigma_{\rm{LO}}=\sum_{u_i,d_j}\int_{\tau}^1\frac{dx}{x}\,f_{u_i}(x,\mu_R)f_{d_j}\left(\tau/x,\mu_R\right)\,\sigma_{u_id_j}
\ea
with $\sqrt{s}$ the CM energy of the hadron collider, $\tau=m_{1/3}^2/s$, and 
\ba
\sigma_{u_id_j}=\frac{\pi}{3}\frac{|\lambda_{ij}|^2}{s}.
\ea
Here $q_i$ denotes the initial quark, taken to be massless, and $f_{q_i}(x,\mu_R)$ is its parton distribution function (PDF) evaluated at the RG scale $\mu_R$. For simplicity we will limit our study to the case $\lambda_{13,23}=0$, so the sum in~(\ref{LO}) extends over the couples $(u_id_j)=(us,su,cd,dc)$ and conjugates.

\begin{figure}
\begin{center}
\includegraphics[width=4.0in]{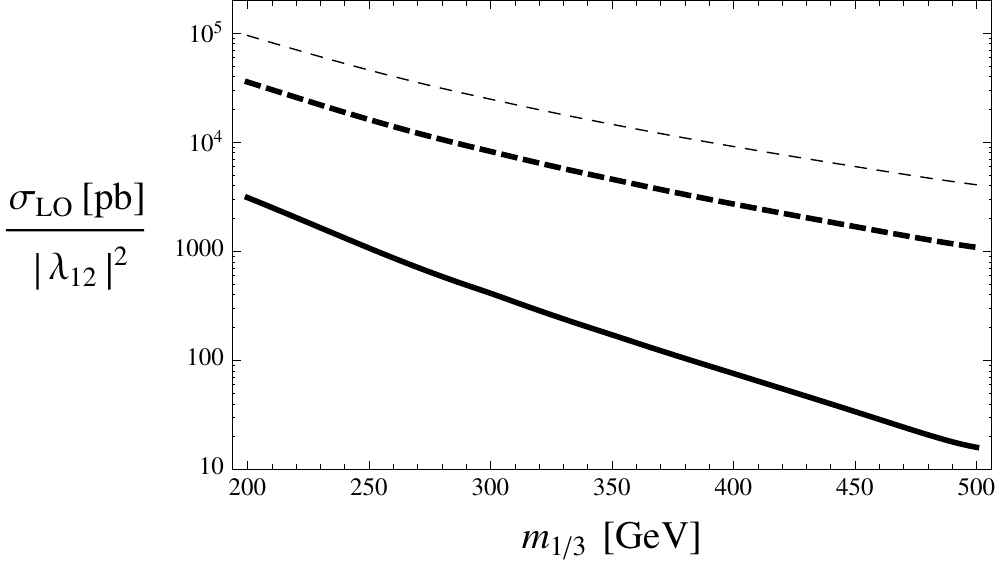}
\caption{\small LO production cross section of $\phi_{1/3}$ at the Tevatron ($\sqrt{s}=1.96$ TeV, see solid thick line) and the LHC ($\sqrt{s}=7/14$ TeV, see dashed thick/dashed thin line) as a function of $m_{1/3}$. The dependence on $|\lambda_{12}|$ has been factorized and the contribution of $\lambda_{13,23}$ neglected. 
\label{prod}}
\end{center}
\end{figure}

The cross section $\sigma_{\rm{LO}}/|\lambda_{ij}|^2$ is shown in fig.~\ref{prod} as a function of $m_{1/3}$ for the Tevatron ($\sqrt{s}=1.96$ TeV, solid thick line) and LHC ($\sqrt{s}=7/14$ TeV, dashed thick/dashed thin line) experiments. We used~(\ref{LO}) and the PDFs of~\cite{MSTW} furnished on a {\tt{Mathematica}} interface at~\cite{MSTW1}. The PDFs have been evaluated at the renormalization scale $\mu_R=m_{1/3}$. We also checked the consistency of the result using {\tt MadGraph/MadEvent v4}~\cite{MG}. The two approaches agree up to at most a $10\%$ discrepancy in the mass range $200$ GeV $<m_{1/3}<500$ GeV. The inclusion of NLO effects in~(\ref{LO}) was estimated to amount to an almost flat $K$-factor of order $\sim1.3$ in~\cite{Kfactor}. Here we focus on the LO analysis.

From fig.~\ref{prod} we learn that, if $\lambda_{12}$ is within the bound shown in~(\ref{generic}), $\sigma_{\rm LO}=O(1)$ pb at the Tevatron are obtained for relatively light masses, see table~\ref{tab}. For example, fixing $m_{1/3}=300$ GeV we find that $\sigma_{\rm LO}\approx4$ pb for $|\lambda_{12}|=0.1$ whereas $\sigma_{\rm LO}\approx1$ pb for $|\lambda_{12}|=0.05$. At the LHC the cross section increases significantly, and for $m_{1/3}=300$ GeV we now have $\sigma_{\rm LO}\approx82$ pb for $|\lambda_{12}|=0.1$ or $\sigma_{\rm LO}\approx21$ pb for $|\lambda_{12}|=0.05$, see table~\ref{tabLHC}. 

\begin{table}
\begin{center}
\begin{tabular}{|c||c|c|}
\hline
$m_{1/3}$ [GeV] & $|\lambda_{12}|=0.05$ & $|\lambda_{12}|=0.10$  \\ 
\hline
$200$  &$7.8$ pb & $31.1$ pb \\ 
 $250$  & $2.7$ pb & $10.7$ pb \\  
 $300$  &$1.0$ pb & $4.1$ pb  \\  
 $350$  &$0.4$ pb & $1.7$ pb  \\
 $400$  &$0.2$ pb & $0.8$ pb  \\  
  $450$  &$0.1$ pb & $0.3$ pb  \\  
 $500$  &$0.04$ pb & $0.2$ pb  \\  
\hline
\end{tabular}
\caption{LO cross section $\sigma(p\bar p\rightarrow\phi_{1/3})$ for resonant $\phi_{1/3}$ production at the Tevatron for a few benchmark points, see also fig.~\ref{prod}. 
\label{tab}}
\end{center}
\end{table}

\begin{table}
\begin{center}
\begin{tabular}{|c||c|c|}
\hline
$m_{1/3}$ [GeV] & $|\lambda_{12}|=0.05$ & $|\lambda_{12}|=0.10$  \\ 
\hline
$200$  &$89.2$ pb & $356.7$ pb \\ 
 $250$  & $40.3$ pb & $161.4$ pb \\  
 $300$  &$20.6$ pb & $82.5$ pb \\  
 $350$  &$11.5$ pb & $45.9$ pb \\
 $400$  &$6.8$ pb & $27.2$ pb \\ 
 $450$  &$4.2$ pb & $16.9$ pb \\ 
 $500$  &$2.7$ pb & $10.9$ pb \\  
\hline
\end{tabular}
\caption{LO cross section $\sigma(p\bar p\rightarrow\phi_{1/3})$ for resonant $\phi_{1/3}$ production at the LHC ($\sqrt{s}=7$ TeV) for a few benchmark points, see also fig.~\ref{prod}. 
\label{tabLHC}}
\end{center}
\end{table}

\begin{figure}
\begin{center}
\includegraphics[width=4.0in]{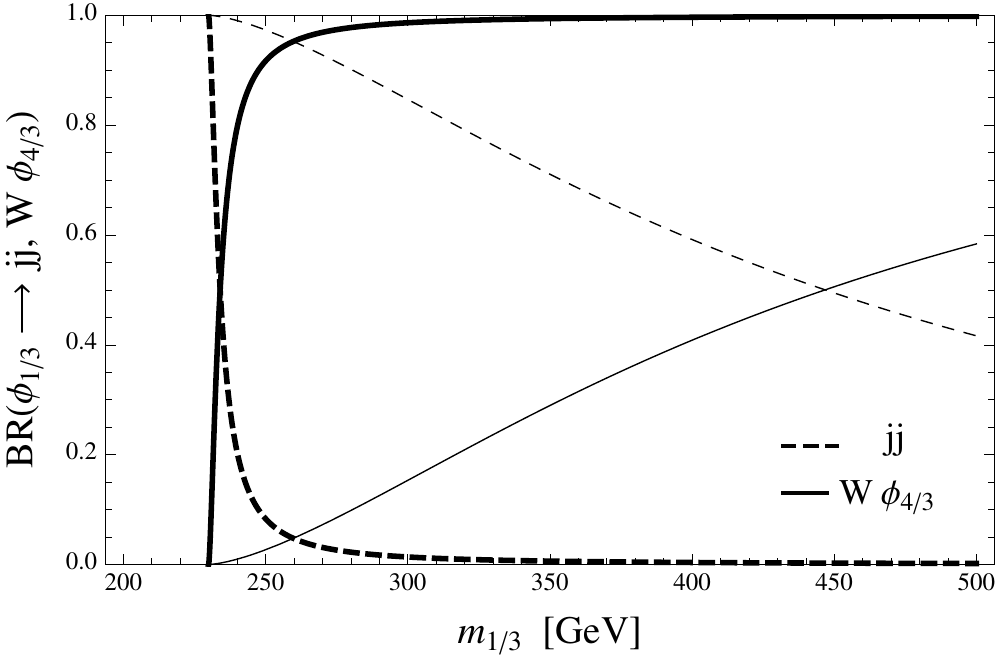}
\caption{\small This plot shows $BR(\phi_{1/3}\rightarrow jj,\phi_{4/3}W)$ as a function of $m_{1/3}$ for $|\lambda_{ij}|=0.05/1$ (see thick/thin lines) and $m_{4/3}=150$ GeV. 
\label{BRfig}}
\end{center}
\end{figure}

Once the scalar (here the $\phi_{1/3}$ component) is produced it decays into di-jets and, if kinematically allowed, into its isospin partners in association with a $W$, here either $\phi_{4/3}W$ or $\phi_{2/3}W$ depending on whether $\Delta m>m_W$ or $\Delta m<-m_W$, see eq.~(\ref{hierarchy}) and the discussion below it. Focussing for definiteness on the case $\Delta m>m_W$, the latter rate is given by:
\ba
\Gamma(\phi_{1/3}\rightarrow\phi_{4/3}W)=\frac{g^2}{16\pi}\frac{m_{\phi_{1/3}}^3}{m_W^2}\,{\cal F}^{3/2}\left(\frac{m_W}{m_{\phi_{1/3}}},\frac{m_{\phi_{4/3}}}{m_{\phi_{1/3}}}\right),
\ea
with ${\cal F}(x,y)=1-2(x^2+y^2)+(x^2-y^2)^2$. 
Assuming for simplicity that only one coupling $\lambda_{ij}$ is relevant, and neglecting the $u,d,s,c$ quark masses, we also find
\ba
\Gamma(\phi_{1/3}\rightarrow jj)=\frac{|\lambda_{ij}|^2}{2\pi}\,m_{\phi_{1/3}}.
\ea
The coupling to the $W^\pm$ is enhanced compared to the case of weak doublets. Meanwhile, when $\lambda_{ij}$ is generic the bounds shown in fig.~\ref{FCNC} apply. As a result, the decay into $\phi_{4/3}W$ quickly saturates the total rate as soon as the mass $m_{1/3}$ exceeds the $m_W+m_{4/3}$ threshold. This is illustrated in fig.~\ref{BRfig} for $|\lambda_{ij}|=0.05$ (thick lines). In this regime fig.~\ref{prod} provides an accurate estimate of the production cross section $\sigma(p\bar p/pp\rightarrow\phi_{4/3}W)$ as soon as $m_{1/3}>m_W+m_{4/3}$ (i.e. $\Delta m>m_W$). In the case of inverted hierarchy, i.e. for $\Delta m<0$, and again for generic couplings, fig.~\ref{prod} provides an accurate estimate of $\sigma(p\bar p/pp\rightarrow\phi_{2/3}W)$ as soon as $m_{1/3}>m_W+m_{2/3}$ (i.e. $\Delta m<-m_W$).

As a reference, in fig.~\ref{BRfig} we also show the branching ratios $BR(\phi_{1/3}\rightarrow jj,\phi_{4/3}W)$ in the case $|\lambda_{ij}|=1$ of nongeneric couplings (thin lines), see section~\ref{hierarchical}.

Comparable production rates to those shown in fig.~\ref{prod} are expected for $\phi_{2/3}$ and $\phi_{4/3}$. 
For $\Delta m>m_W$ the $\phi_{2/3}$ would decay into $W^+\phi^*_{1/3}\rightarrow W^+W^+\phi^*_{4/3}$ leading to a characteristic $2l2\nu jj$ final state with a same-sign dilepton. On the other hand, in the case of resonant $\phi_{4/3}$ production the decay will be entirely into jets. For $\Delta m<-m_W$ the pattern is reversed: the component $\phi_{2/3}$ would be the lightest element of the triplet, with a branching fraction dominantly into jets (now also $b$-jets), while $\phi_{4/3}$ would now be the heavier decaying into same-sign dileptons.

The detection of the triplet at the Tevatron and the LHC is facilitated in the case the decays into light partners are kinematically open. If this condition is not met the decay products would be entirely jets and hence overwhelmed by the QCD background, especially for weak scale scalars. In this pessimistic scenario one can still hope the couplings $\lambda_{13,23}$ to the third quark generation are unsuppressed, in which case the production of the color $\&$ weak triplet would be characterized by its flavor-violating decays into top and bottom quarks, such as $\phi_{4/3}\rightarrow ut,ct$, or $\phi_{1/3}\rightarrow dt,st,bu,bc$, and $\phi_{2/3}\rightarrow d^*b^*,s^*b^*$. A dedicated analysis of the significance of same-sign dilepton decays and flavor-violating decays at hadron colliders is beyond the purpose of this paper.

We conclude this section mentioning that the bounds imposed by the di-jet searches at the UA2~\cite{UA2} and Tevatron~\cite{dijets1995}\cite{dijets2003}\cite{dijets2008} experiments, as well as the ATLAS collaboration with an integrated luminosity of $315$ nb$^{-1}$~\cite{ATLASdijets} and $36$ pb$^{-1}$~\cite{ATLASdijets2}, do not constrain the color $\&$ weak triplet model further, even if we conservatively assume that the resonance has $BR(\phi\rightarrow jj)=1$. 

One can find additional details on the phenomenology of colored scalars in \cite{MW}\cite{Gripaios}\cite{list}\cite{Kfactor}\cite{s}\cite{FB}\cite{Kamenik2}.

\subsection{Higgs physics}

The Higgs physics can be significantly affected by new colored particles. Color $\&$ weak triplets can for instance modify the processes $gg,\gamma\gamma\rightarrow h$. This effect is described by higher dimensional operators $\sim\kappa vhF_{\mu\nu}F_{\mu\nu}/(16\pi^2m_\phi^2)$, with $F_{\mu\nu}$ the (noncanonically normalized) field strength of either the gluon or the photon. These operators can constructively or destructively interfere with the SM depending on the sign of the effective triple coupling $\kappa$, and can easily lead to sizable corrections of the SM expectations when $m_\phi$ is of weak scale magnitude~\cite{MWh}.

One might worry that a sizable correction to the Higgs production rate could be in conflict with the recent data from the ATLAS collaboration~\cite{atlas}. However, these bounds are not constraining if the Higgs is sufficiently heavy or/and if the fraction $BR(h\rightarrow WW)$ is suppressed~\cite{Gunion}. Fortunately, in our model both effects may be present.

First, we saw that EW precision measurements tend to favor a Higgs boson heavier than in the minimal SM as soon as a tiny mass splitting among the elements of the triplet is turned on. Higgs masses above $200$ GeV should be sufficient to make our model compatible with direct Higgs searches for a generic $\kappa$~\cite{Gunion}.

Second, when $m_h>2m_\phi$ the branching ratio for $h\to\phi\phi$ can potentially compete with $BR(h\rightarrow WW)$. Specifically, when the Higgs mass is above the diboson threshold we find
\ba\label{ratio}
\frac{BR(h\rightarrow\phi\phi)}{BR(h\rightarrow WW)}=6\kappa^2\frac{v^4}{m_h^4}\frac{\sqrt{1-4x_\phi}}{\sqrt{1-4x_W}\left(1-4x_W+12x_W^2\right)},
\ea
with $\kappa=\kappa_1,\kappa_1-\kappa_2$, and $\kappa_1+\kappa_2$ for $\phi=\phi_{1/3},\phi_{2/3}$, and $\phi_{4/3}$ respectively ($\kappa_{1,2}$ are defined in~(\ref{potential})), whereas $x_{\phi,W}=m_{\phi,W}^2/m_h^2$. 
The expression~(\ref{ratio}) is a strong function of the Higgs mass $m_h$, which reveals a maximum at $m_h\gtrsim2m_\phi$. When $m_h\gg v$, the $h\rightarrow WW$ mode always dominates. For $m_h\lesssim v$, however, the $h\rightarrow\phi\phi$ channel has a rate typically comparable to $h\rightarrow WW$, ranging between $25\%\lesssim BR(h\rightarrow\phi\phi)\lesssim 55\%$ for $\kappa=1$, such that when $\kappa\gtrsim1$ it can easily dominate. The latter possibility is especially interesting because in such a scenario the signature of a heavy Higgs boson would 
become $h\rightarrow4j$.

\section{Applications: the Tevatron anomalies}
\label{anomaly}

In this section we turn to a study of the recent Tevatron anomalies~\cite{FBAD0}\cite{FBACDF}\cite{bump}\cite{like-sign}. 

The top FBA has already been studied in similar models in~\cite{FB}, so we refer the reader to those papers for more details. Our models for the CDF $Wjj$ anomaly and the dimuon asymmetry are instead original. A characteristic feature of our model for the CDF $Wjj$ anomaly is the prediction of no $b$-quark jets in the excess region $120$ GeV $<M_{jj}<160$ GeV. The model for the dimuon asymmetry is characterized by the prediction $h_d\ll h_s$ (see eq. (\ref{hshd}) for a definition of these parameters), and shares some analogies with the ``maximally flavor-violating" model of~\cite{SZ}.

We will see that while the CDF $Wjj$ excess and the dimuon asymmetry can be explained within the color $\&$ weak triplet framework with generic Yukawa matrices $\lambda_{ij}$, the observed top FBA can only be explained provided one excepts the existence of the hierarchical relation $|\lambda_{23}|\ll|\lambda_{12}|\ll|\lambda_{13}|\sim1$. Once the hypothesis of hierarchical couplings $\lambda_{ij}$ is excepted, one finds that sizable corrections to the physics of the $B_s$ meson -- but not to $B_d$ -- can be induced in a way compatible with all other bounds. Remarkably, these conditions are required to account for the observed dimuon charge asymmetry in $b$ decay~\cite{like-sign}\cite{Papucci} and the very recent measurement of $BR(B_s\to\mu^+\mu^-)$~\cite{mumu}.

\subsection{The CDF $Wjj$ excess}

The CDF collaboration reported an excess in the invariant mass distribution of di-jets produced in association with a $W^\pm$~\cite{bump}. It is not yet clear if the CDF anomaly is due to NP or to SM background~\cite{D0}. In this subsection we assume that the $Wjj$ excess is due to a beyond the SM di-jet resonance of mass $\sim150$ GeV~\cite{bump}.

The color $\&$ weak triplet model has been previously discussed in the context of the CDF anomaly in~\cite{s}. In that work the components of the triplet were taken to be degenerate in mass, so the excess in the $120$ GeV $<M_{jj}<160$ GeV invariant mass distribution was attributed to any of the $\phi_{1/3, 2/3,4/3}$. Furthermore, the Yukawa matrix $\lambda_{ij}$ was assumed to be hierarchical in order to generate an $O(1)$ pb cross section for the $t$-channel process $p\overline{p}\rightarrow\phi W$. 

In the present paper the mass splitting between the elements of the triplet plays a crucial role: we will see that if $\phi_{4/3}$ happens to be the lightest scalar it can naturally play the role of the resonance observed by CDF and that $\sigma(p\overline{p}\rightarrow\phi_{4/3} W)=O(1)$ pb can be attained while keeping $\lambda_{12}$ within the range~(\ref{generic}).

We take $\phi_{4/3}$ to be the lightest component of the triplet, with a mass $m_{4/3}\sim150$ GeV. Under this hypothesis, and working within a perturbative description in which the relation~(\ref{tree}) is accurate, one finds that $m_{2/3}>m_{1/3}>m_{4/3}$. Here we assume that the mass splitting is of order the weak scale, $\Delta m>m_W$, so that:
\ba\label{spectrum}
m_{4/3}+m_W<m_{1/3}<m_{2/3}-m_W.
\ea
We discussed how it is possible to make the pattern~(\ref{spectrum}) compatible with the EW bounds in section~\ref{EWbounds}.

We saw in the previous section that the mass pattern~(\ref{spectrum}) is an ideal condition for detection. We now show that, given this hypothesis on the spectrum, 
the resonance $\phi_{4/3}$ naturally satisfies three crucial features~\cite{bump}: (1) it is produced with a rather large rate and (2) preferentially in association with a $W$, and finally (3) it does decay dominantly into light jets.


The absence of excess of $b$-quarks in the $p\overline{p}\rightarrow\phi_{4/3}W\rightarrow jjW$ final state is easy to understand, for by charge conservation $\phi_{4/3}$ will decay isotropically into up-type quarks: $\phi_{4/3}\rightarrow u_iu_j$ with $BR(\phi_{4/3}\rightarrow jj)\approx1$.

The suppression of the analogous $p\overline{p}\rightarrow Zjj$ channel is a consequence of gauge invariance as well as dynamics. Indeed, the scalar $\phi_{4/3}$ can be produced in association with a $W^\pm$ only by $qq$ scattering, where the initial partons have total electric charge $Q=\pm1/3$, i.e. $u_id_j\rightarrow W^-\phi_{4/3}$. The production in association with a $Z^0$ is instead of the type $u_iu_j\rightarrow Z^0\phi_{4/3}$. Recalling that the Yukawa coupling $\lambda_{ij}$ is antisymmetric in the family index, we see that the production in association with $W$ occurs dominantly via initial $us, cd$ pairs, while the production in association with $Z$ via initial $uc$. The $Zjj$ channel is thus suppressed compared to $W jj$ by the c-quark PDF and a lower multiplicity. This suppression is $O(1)$ if the processes $p\overline{p}\rightarrow\phi_{4/3}W\rightarrow jjl\nu$ occur dominantly via a $t$-channel exchange~\cite{s}, while it is estimated to be $\gtrsim O(10)$ if~(\ref{spectrum}) holds, in which case the $W\phi_{4/3}$ final state is enhanced by the resonant production $u_id_j\rightarrow\phi_{1/3}\rightarrow W\phi_{4/3}$, see also~\cite{carena}. No such enhancement is possible for the analogous $Z\phi_{4/3}$ final state. 

Resonant $\phi_{1/3}$ production has been studied in section~\ref{resprod}, where it was argued that an $O(1)$ pb cross section for $p\overline{p}\rightarrow\phi_{1/3}\rightarrow\phi_{4/3} W\rightarrow jjW$ is consistent with the experimental bounds discussed in section~\ref{model}.

We thus see that:
\begin{itemize}
\item the color $\&$ weak triplet model can explain the $Wjj$ excess reported by CDF~\cite{bump} with generic, nonhierarchical couplings and a mass splitting $\Delta m\gtrsim m_W$.
\end{itemize}
Characterizing features of our model for the CDF anomaly are: (i) the absence of $b$-quark jets in the excess region, and a suppression of at least an $O(10)$ factor of the associated $p\overline{p}\rightarrow Zjj$ process; (ii) a peak in the $Wjj$ invariant mass distribution at $M_{jjW}\approx m_{1/3}$. The analysis of section~\ref{resprod} suggests that $250$ GeV $\lesssim m_{1/3}\lesssim450$ GeV would be required to have $\sigma(p\overline{p}\rightarrow\phi_{4/3} W)\sim0.1-11$ pb, see the benchmark points in table~\ref{tab}. The value $m_{1/3}\sim300$ GeV might be consistent with the results presented at~\cite{Mother}; (iii) the prediction of a characteristic pattern $p\overline{p}\rightarrow\phi_{2/3}\rightarrow W\phi_{1/3}\rightarrow WWjj$ at a higher invariant mass $M_{jjWW}\approx 320-620$ GeV (here we took $m_{1/3}=250-450$ GeV from table~\ref{tab} and used~(\ref{tree})). We also expect additional fields not far from the weak scale to make the large mass splitting $|\Delta m|\gtrsim m_W$ compatible with the EW data (see section~\ref{EWbounds}). 

\subsection{The CDF $Wjj$ excess + the top FBA}

Colored diquarks can account for the observed $t\overline{t}$ forward-backward asymmetry (FBA)~\cite{FB}. 
Here we will analyze under which conditions a simultaneous explanation of the top FBA and the CDF $Wjj$ excess is possible.

The aim of any model for the top FBA is to generate a large NP contribution to the FBA, but not to the total $t\overline{t}$ production cross section. Within the color $\&$ weak triplet framework, these processes are dominantly altered by the $u$-channel exchange of $\phi_{4/3}$ and $\phi_{1/3}$. In either case the relevant coupling is $\lambda_{13}$. 

In~\cite{FB} 
it was shown that a NP contribution to the FBA having the correct sign and magnitude is attained only for masses $m_\phi\gtrsim300$ GeV. If we are assume $m_{4/3}\sim150$ GeV, as required to fit the CDF excess, we see that the field playing the dominant role in correcting the SM top FBA will be $\phi_{1/3}$. Because $\phi_{4/3}$ is expected to be subdominant, we conservatively neglect its contribution in the following.

The coupling $\lambda_{13}$ controlling the NP corrections to $d\sigma(p\overline{p}\to t\overline{t})/d\cos\theta$ should be rather large~\footnote{This fact implies that single-top production is enhanced compared to the SM~\cite{Strassler}. To avoid conflict with data the NP contribution to the top FBA should be smaller than currently expected. This is certainly the case if the SM prediction has been underestimated, as suggested for example in~\cite{FBweak}.}. Indeed, it was found that both FBA and $tt^*$ production can be made consistent with data if the Yukawa coupling $\lambda_{13}$ satisfies an approximately linear relation with the mass $m_{\phi}$. Allowing at most a one-standard deviation from the observed rates, 
this relation reads~\cite{FB}:
\ba\label{linear}
|\lambda_{13}|=\frac{m_{\phi}}{830~{\rm GeV}}+0.5.
\ea
Strictly speaking, eq.~(\ref{linear}) has been derived in~\cite{FB} when the dominant production process is $uu^*\rightarrow tt^*$. If $m_{4/3}\sim150$ GeV and $m_{1/3}\gtrsim300$ GeV, however, the dominant reaction is actually $dd^*\to tt^*$. 
We will nevertheless use~(\ref{linear}) as a rough estimate of the magnitude of $|\lambda_{13}|$ required to explain the measured top FBA. 

For the values $m_{1/3}\sim250-300$ GeV favored by the CDF $Wjj$ excess we find $|\lambda_{13}|\sim0.8-1.0$. In section~\ref{hierarchical} we will see that such large values of $|\lambda_{13}|$ can be compatible with meson oscillation provided $|\lambda_{12}|,|\lambda_{23}|\ll1$. Now, because $BR(\phi_{1/3}\to ub,dt)\sim1$ when $|\lambda_{13}|\sim1$, see fig.~\ref{BRfig}, it is clear that the rate for the process $p\overline{p}\to W\phi_{4/3}$ would be too small to account for the results reported in~\cite{bump}. We thus conclude that:
\begin{itemize}
\item the color $\&$ weak triplet model can explain either the top FBA~\cite{FBAD0}\cite{FBACDF} or the CDF $Wjj$ excess~\cite{bump}, but not both anomalies simultaneously. (The top FBA is explained with $|\lambda_{13}|\sim1$ and $m_\phi\gtrsim300$ GeV, whereas $\Delta m$ is not constrained~\cite{FB}.)
\end{itemize}

\subsection{The top FBA + the dimuon asymmetry}
\label{hierarchical}

In this subsection we discuss under which conditions the value $|\lambda_{13}|\sim1$ required to explain the top FBA is compatible with the constraints on flavor violation. 

First, note that NP contributions to $D-\overline{D}$ are governed by the Wilson coefficient $C_D^1\approx(\lambda_{13}\lambda_{23}^*)^2/(4\pi^2m_{4/3}^2)$, so that to avoid too large a correction to $D$-meson oscillation we will impose $|\lambda_{23}|<10^{-3}$ when $m_\phi$ is at the weak scale~\cite{UTFit} (see also~\cite{Kamenik2}). 

\begin{figure}
\begin{center}
\includegraphics[width=4.5in]{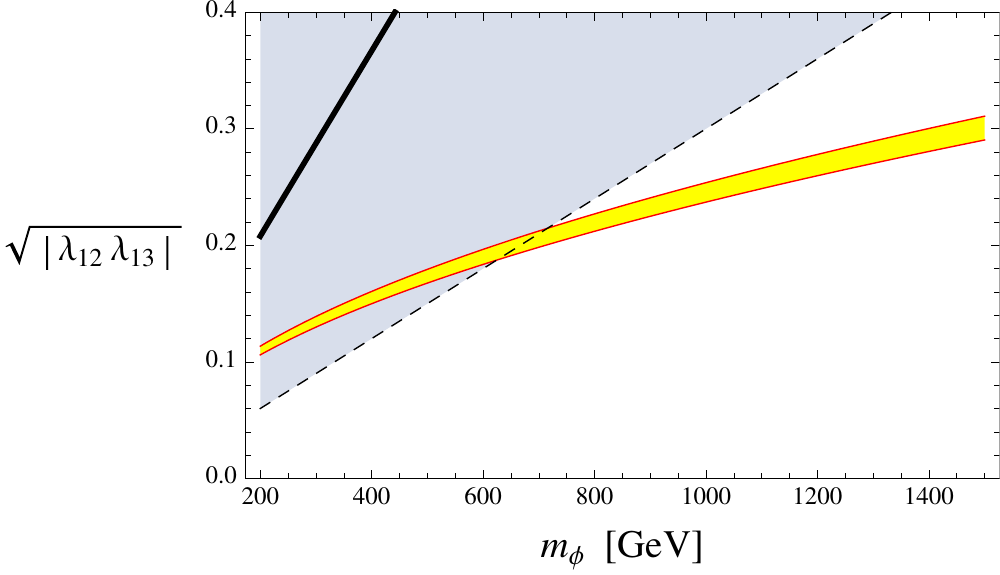}
\caption{\small (Hierarchical Yukawa couplings) Values of $\sqrt{|\lambda_{12}\lambda_{13}^*|}$ excluded by $\Delta F=1,2$ bounds (grey area) and the region preferred by the dimuon asymmetry~\cite{like-sign} (yellow strip). The constraint from $D-\overline{D}$ is satisfied by requiring $|\lambda_{23}|<10^{-3}$. The solid, thick black line is the bound from $K-\overline{K}$. The dashed thin line is the $20\%$ bound from the $\Delta F=1$ processes shown in fig.~\ref{FCNC}. The SM prediction for $B_d-\overline{B_d}$ and $K\to\pi\nu\overline{\nu}$ are very accurate in the allowed parameter space (white area). The yellow strip between the solid, red lines corresponds to the $2\sigma$ region for the dimuon asymmetry, which roughly reads $1.8<h_s<2.1$~\cite{Papucci}. \label{b}}
\end{center}
\end{figure}

Now, if either $\lambda_{13}$ or $\lambda_{23}$ are smaller than $O(10^{-2}|\lambda_{12}|)$  the approximation used in our study of $K-\overline{K}$ mixing and leading to~(\ref{C1app}) breaks down. We here discuss how the bounds of fig.~\ref{FCNC} change in the limit in which $\lambda_{23}\to0$, but analogous results hold if $\lambda_{13}$ is small. When $|\lambda_{23}V_{33}|\ll|\lambda_{12}V_{13}|$ and $|\lambda_{12}|\lesssim|\lambda_{13}|$ one sees that $\sum_i\xi_i\ll\xi_3$ and the $\phi_{2/3}$ contribution in~(\ref{C1}) becomes negligible. We hence have:
\ba\label{CKh}
C_K^1(m_{1/3})\approx\frac{\xi_3^2}{16\pi^2}\left[\frac{{\cal I}(0,0)+{\cal I}(x^u_3,x^u_3)-2{\cal I}(x^u_3,0)}{m_{1/3}^2}\right],
\ea
where $|\xi_3|\approx|V_{13}\lambda_{12}V^*_{33}\lambda_{13}^*|$. The bound~(\ref{boundKK}) on this coefficient is shown in fig.~\ref{b} (thick, solid line). Note that the bound is \emph{weaker} than that imposed by $\Delta F=1$ violation~(\ref{BtoS}).

The constraints from $B_q-\overline{B}_q$ are also affected by the hierarchical relation among the couplings $\lambda_{ij}$. In general, while the $b\leftrightarrow d$ transitions are strongly suppressed for $\lambda_{23}\to0$, the $b\leftrightarrow s$ transitions are not.

Indeed, if $|\lambda_{23}|\ll1$ the Wilson coefficient $C_{B_d}^1$ controlling $B_d-\overline{B}_d$ is basically the same as~(\ref{CKh}), but with $\xi_3\approx V_{13}V_{33}|\lambda_{13}|^2$. This leads to a negligible correction of the SM prediction even when $|\lambda_{13}|=O(1)$. In standard notation the dispersive part of the $B_q-\overline{B}_q$ amplitude is written as
\ba\label{hshd}
M_{12}^q=M_{12}^{q,\,\rm SM}\left(1+h_qe^{2i\sigma_q}\right),
\ea
and our result reads $h_d\ll1$.

The NP contribution to the Wilson coefficient $C_{B_s}^1$ of the operator $(\overline{s}\gamma^\mu b)(\overline{s}\gamma^\mu b)$ controlling $B_s-\overline{B}_s$ is well approximated by 
\ba
C_{B_s}^1\approx\frac{\left(\lambda_{21}^*\lambda_{31}\right)^2}{4\pi^2}\left[\frac{{\cal I}(x^u_3,x^u_3)}{4m_{1/3}^2}+\frac{{\cal I}(0,0)}{m_{2/3}^2}\right],
\ea
irrespective of whether $\lambda_{ij}$ is hierarchical or not. By adjusting the coupling $\lambda_{12}$ (while still respecting the other bounds) the NP contribution to $B_s$ oscillation can easily be made comparable to the SM in the regime $|\lambda_{13}|\sim1$ favored by the top FBA~\cite{FBAD0}\cite{FBACDF}. In standard notation this result reads $h_s=O(1)$. 

Within two sigma deviations, the measured value of the like-sign dimuon charge asymmetry in semileptonic $b$ decay can be explained by a NP model with negligible corrections to $B_d-\overline{B_d}$ if the contribution to $B_s-\overline{B_s}$ is such that $h_s\sim1.8-2.1$~\cite{Papucci}\cite{Nir} (the results of~\cite{Papucci} also tell us the phase of $\lambda_{21}^*\lambda_{31}$, which we assumed generic throughout our analysis). We show in  fig.~\ref{b} that in the color $\&$ weak triplet model this condition is satisfied for masses $m_\phi\gtrsim700$ GeV and couplings $|\lambda_{12}|\gtrsim0.04/|\lambda_{13}|$.

We thus conclude that:
\begin{itemize}
\item the color $\&$ weak triplet model can simultaneously explain the top FBA~\cite{FBAD0}\cite{FBACDF} and the like-sign dimuon charge asymmetry in semileptonic $b$ decay~\cite{like-sign}, without spoiling the success of the SM in explaining $\Delta F=1$ processes and $K,D, B_d$-meson oscillation. This can be achieved provided the scalar has weak scale masses ($\Delta m$ is not constrained) and hierarchical Yukawa matrices satisfying $|\lambda_{23}|\ll|\lambda_{12}|\ll|\lambda_{13}|\sim1$.
\end{itemize}

CDF has also reported a measurement of the branching fraction for $B_s,B_d\to\mu^+\mu^-$ and found that, while for $B_d\to\mu^+\mu^-$ the SM prediction is accurate, $BR(B_s\to\mu^+\mu^-)$ is a factor $O(1)$ larger than expected~\cite{mumu}. We here point out that this enhancement is generally present in our model for the top FBA and the like-sign dimuon charge asymmetry. 
A precise estimate of this effect would lead to an additional line in fig.~\ref{b} and possibly determine both $\sqrt{|\lambda_{12}\lambda_{13}|}$ and $m_\phi$.

\subsection{The dimuon asymmetry + the $Wjj$ excess}

One can show with arguments similar to those of the previous subsections that:
\begin{itemize}
\item the color $\&$ weak triplet model can simultaneously explain the CDF $Wjj$ anomaly~\cite{bump} and the like-sign dimuon charge asymmetry in semileptonic $b$ decay~\cite{like-sign}. This can be achieved provided $m_{4/3}\sim150$ GeV and $\Delta m\gtrsim m_W$, and for generic Yukawa matrices  satisfying $|\lambda_{23}|<|\lambda_{13}|<|\lambda_{12}|\sim0.1$.
\end{itemize}

\section{Conclusions}

\label{conclusions}

In this paper we discussed in some details the phenomenology of color $\&$ weak triplet scalars with hypercharge $Y=-1/3$. 

We found that SM gauge invariance forces the renormalizable Yukawa couplings to di-quarks to be antisymmetric. This in turn implies that $\Delta F=2$ processes first arise at the one-loop level, and that the resulting NP theory is ``maximally flavor-violating", as suggested by the recent Tevatron data. At tree-level, the triplet only contributes to non-leptonic $\Delta F=1$ heavy meson decays, and generic, non-hierarchical, sizable couplings $\sim0.05-0.1$ to the SM quarks 
are found to be consistent with data for scalars with weak scale masses. 

After electro-weak symmetry breaking the elements of the weak triplet split in mass according to their isospin. We analyzed the impact of the mass splitting on the electroweak observables and showed that -- for nonvanishing splitting -- 
Higgs masses above the LEP bound are favored by data.

The triplet can substantially alter the SM Higgs production at hadron colliders. 
Furthermore, for Higgs masses above the $WW$ and $\phi\phi$ thresholds, $h\rightarrow\phi\phi\rightarrow4j$ might in fact represent the dominant decay mode, making the detection of the Higgs boson at the LHC more subtle.

Triplets with weak scale masses also lead to interesting signatures at hadron colliders. 
We estimated the LO cross section for resonant production at the Tevatron and the LHC, and discussed the main decay modes. The cleanest signal of the color $\&$ weak triplet occurs when the mass splitting is of weak scale magnitude. In this case the decay of the heaviest component leads to same-sign dilepton final states, whereas the intermediate component would dominantly decay into the lightest component in association with a $W$. If these latter channels are kinematically closed, then the triplet would decay entirely into jets. Whether or not large branching ratios into tops or bottoms are possible depends on the details of the Yukawa matrix. This should be compared to what happens in theories in which minimal flavor violation is imposed, in which case the couplings to the third generation are typically larger.

Our interest on the physics of the color $\&$ weak triplet model was motivated by a number of anomalies reported by the Tevatron groups~\cite{FBAD0}\cite{FBACDF}\cite{bump}\cite{like-sign}. We turned our attention to a possible explanation of these anomalies within the color $\&$ weak triplet model in section~\ref{anomaly}.

A simultaneous explanation of the like-sign dimuon charge asymmetry in semileptonic $b$ decay~\cite{like-sign} \emph{and} the top FBA~\cite{FBAD0}\cite{FBACDF} is possible~\footnote{The only paper we are aware of in which the same goal is pursued is~\cite{SZ}.} for weak scale masses $m_\phi$ (and generic $\Delta m$) and provided hierarchical relations among the Yukawa couplings of the triplet are present. Remarkably, a characterizing feature of our model is the prediction of $h_d\ll h_s$, a relation which seems to be suggested by data~\cite{like-sign}\cite{Papucci}. 

The dimuon asymmetry may be accommodated within our model for the CDF $Wjj$ anomaly, as well. However, the CDF $Wjj$ excess~\cite{bump} was shown to be \emph{incompatible} with the top FBA in our minimal framework.


Characterizing properties of our model for the $p\overline{p}\rightarrow Wjj$ CDF anomaly are the prediction of no excess of $b$-quarks in the $Wjj$ final state and a suppression of the analogous $p\overline{p}\rightarrow Zjj$ channel. The flavor problem typically required to explain these two latter features has been converted into a \emph{spectrum problem} within the context of the color $\&$ weak triplet model. 

\acknowledgments
We thank Vincenzo Cirigliano, Michael Graesser, and Ian Shoemaker for discussions.
This work has been supported by the U.S. Department of Energy at Los 
Alamos National Laboratory under Contract No. DE-AC52-06NA25396. The preprint number for this manuscript is LA-UR-11-10996.


 \end{document}